\documentclass[10pt,
twocolumn,
aps,prd,nofootinbib]
{revtex4-2}
\usepackage[utf8]{inputenc}
\usepackage{bm} 
\usepackage{physics} 
\usepackage{dcolumn} 
\usepackage{amsfonts,amsmath,amssymb} 
\usepackage{graphicx,xcolor} 
\usepackage{hyperref,cleveref}
\usepackage{siunitx}
\usepackage{etoolbox}
\apptocmd{\sloppy}{\hbadness 10000\relax}{}{}
\begin{document}

\title{How the Universe postpones the evaporation and curtails the quantum spreading of black holes}

\author{Quinn Taylor}
\email{quinn.taylor2@case.edu}
\affiliation{Physics Department/CERCA/ISO, Case Western Reserve University, Cleveland, Ohio 44106-7079, USA}

\author{Glenn D. Starkman}
\affiliation{Physics Department/CERCA/ISO, Case Western Reserve University, Cleveland, Ohio 44106-7079, USA\break
Astrophysics, Imperial College London, Blackett Laboratory, 
Prince Consort Road, London, SW7 2AZ, UK}

\date{\today}

\Crefname{equation}{Eq.}{Eqs.}
\Crefname{Figure}{Fig.}{Figs.}

\begin{abstract}
    Black holes are expected to evaporate through the process of Hawking radiation.
    This process is expected to cause the uncertainty in a 
    black hole's position to grow to $\sim M^2/M_{Pl}^3$ over the course of it's lifetime, even as its momentum spreads only by $\sim M_{Pl}$.
    For the black holes that have been observed, which have $M\geq M_\odot$, this greatly exceeds the Hubble volume.
    However, the decay of black holes and their quantum spreading, are delayed in the Universe while the influx of energy into the black holes exceeds their Hawking luminosity.
    We show that for these $M\geq M_\odot$ black holes, their decay outside galaxies and clusters is prevented far longer than it takes the black holes to be dragged well beyond the Hubble horizon, where their eventual decay occurs away from the prying eyes of any observer who has not hitched a ride with them.
    Meanwhile, black holes in an observer's galaxy or cluster are themselves prevented from decaying long past the extinction of the last stars, and at least until their galaxy/cluster is swept clean of dark matter, in
    $\gg 10^{25}$y.
    Even then, if the black holes become unbound they are dragged beyond the Hubble radius before undergoing significant decay;
    if not, they remain in bound orbits,  spreading at most over a subvolume of the galaxy/cluster, and long localized by scatterings to much smaller volumes.
\end{abstract}
    
\maketitle

\section{Introduction}

It has long been expected \cite{Hawking} that although black holes (BHs) are classically eternal, they will evaporate through the quantum process of Hawking radiation.
Page \cite{page_is_1980} argued that this quantum evaporation will result in the substantial  spreading of the black-hole wave function, so that, by the time  a black hole of initial mass $M_i$ has evaporated away a fraction $f$ of its mass (so that $M=(1-f)M_i$), it will have a positional uncertainty
\footnote{    
    This is obtained from equation 2.4a of \cite{Flanagan:2021svq} in the limit where $(1-f)>\sqrt{M_{Pl}/M_i}$.
    Both \cite{Flanagan:2021svq} and  \cite{Nomura2013} explore general values of $f$ in refining the calculation of \cite{page_is_1980}.
}
\begin{flalign}
    \label{eq:Deltax}
    \Delta x 
    &\simeq f^{3/2}  
    \left(\frac{M_i}{M_
    {Pl}}\right)^2
    \ell_{Pl}\\
    &\simeq f^{3/2}
    \left(\frac{M_i}{M_
    \odot}\right)^2
    1.4\times10^{25}\mathrm{ly}\,,\nonumber
    \intertext{despite a momentum uncertainty of just}
    \Delta p &\simeq f^{1/2}M_{Pl}\,.
\end{flalign}
Here $M_{Pl}\simeq2.2\times10^{-5}$g and $\ell_{Pl}\simeq1.6\times10^{-33}$cm are the Planck mass and length, while the solar mass is $M_\odot\simeq2\times10^{33}\si{g}$.
For the solar-and-greater mass black holes that we actually observe, this positional uncertainty
\eqref{eq:Deltax} is enormous.

Others \cite{Nomura2013}, most recently Flanagan \cite{Flanagan:2021svq}, have expanded on Page's investigation, but always in the idealized case of an isolated black hole. 
In this paper we explore whether such spreading will ever be visible for the black holes we know to exist -- those with masses in excess of a solar mass.

\section{The Cosmic Microwave Background delays BH evaporation}

The black hole decay process takes a very long time for a macroscopic black hole,
\begin{flalign}
    \label{eq:Deltat}
    \Delta t &\simeq f \left(\frac{M_i}{M_{Pl}}\right)^3 15360\pi t_{Pl} \\
    &\simeq f \left(\frac{M_i}{M_
    \odot}\right)^3 6.\times10^{67} \mathrm{y},
\nonumber
\end{flalign}
However, for isolated black holes of stellar mass it is ongoing. 
The time between Hawking particle emissions can be found by taking $f\simeq T_H/M_i$ in \eqref{eq:Deltat},
\begin{flalign}
    \delta t_H &\simeq \left(\frac{M_i}{M_
    \odot}\right) 3.\times10^{-3} \mathrm{s}\,.
\end{flalign}
For stellar mass black holes this is, perhaps surprisingly, short.  
For all but the most massive observed supermassive black holes it is at least shorter than the current age of the universe.

However, real black holes are not isolated, they inhabit a universe filled with matter and radiation. 
It is well-understood that such large black holes are not currently undergoing net Hawking evaporation, nor quantum spreading.
This because the Hawking temperature is much lower than the temperature of the cosmic microwave background (CMB)
\begin{flalign}
    \label{THvsTCMB}
    T_H &= \left(\frac{M_
    \odot}{M_i}\right) 6\times10^{-8}\si{K} \ll T_{CMB} \simeq 2.7\si{K}
\end{flalign}
The CMB blackbody photons infalling onto the black hole outpaces the Hawking radiation that the hole radiates.
The photons also collapse the black-hole wave-function, interrupting the quantum spreading.

While $T_H\ll T_{CMB}$ today, the universe has entered a period of exponential expansion with an e-folding time of 
\begin{equation}
    \tau_{Hubble} \simeq \frac{c}{H_0\sqrt{\Omega_\Lambda}} 
    \simeq 16 \si{Gyr} \,.
\end{equation}
The temperature of the CMB is thus falling exponential with this time constant, and
approximating $T_{CMB}(t)\simeq T_{CMB}(t_0)\exp[-(t-t_0)/t_{Hubble}]$,
$T_{CMB}$ will fall to $T_H(M_i)$ at
\begin{equation}
    \label{eqn:Hawkingonset}
    t_\gamma^{eq}(M_i) \simeq \tau_{Hubble}\left(18 + \ln\left[\frac{M_i}{M_
    \odot}\right]\right)\,.
\end{equation}
According to \eqref{eqn:Hawkingonset},
in under $300$ billion years we would expect to enter a period where the evolution of the least massive black holes we currently observe ($M\simeq M_\odot$) is dominated by their Hawking evaporation. 
Even the most massive black hole detected, TON618 ($M_{TON618}\simeq6.6\times10^{10}M_\odot$ \cite{TON618BHmass}),  would  begin evaporating within $7$ trillion years.  
Black hole wave functions would then begin spreading as long envisaged.

However, there is more in the universe than the CMB to disturb the peace, quiet and quantum coherence of massive black holes.
As we show below, it will be a far far longer time before the black holes we currently observe begin evaporating, and we (or our long-from-now descendants) will never be in a position to observe most of them.

\section{The Cosmic Neutrino Background further delays BH evaporation}

CMB photons are not the only significant sources of energy incident on a real black hole;
the cosmic neutrino background (CNB) has not been directly detected but its existence has been inferred from both detailed analysis of the CMB, and light element abundances.
Although the number density of CNB neutrinos is expected to be somewhat lower than that of CMB photons, with a blackbody temperature of $T_{C\nu B}\simeq (4/11)^{1/3}T_{CMB}\simeq 1.9\si{K}$, the CNB should have a somewhat larger energy density.\footnote{
    This assumes that the neutrino species have no sizable chemical potential.   Their total energy density today could be significantly larger if they do.
} 
This is because observations of neutrino mixing indicate that at least one of the three species of neutrinos has a mass $m>0.05$eV, whereas
$kT_{CMB}/c^2\simeq2.3\times10^{-4}$eV.
Such neutrinos would be non-relativistic, 
$v_\nu^{rms}\simeq T_\nu/m<5\times10^{-3}c$,
resulting in a  Sommerfeld enhancement in the cross-section of these neutrinos ---
$\sigma v \propto v(1+c^2/v^2)$ ---
which increases the capture rate of neutrinos dramatically.
Thus,
\begin{flalign}
    \label{eq:tnuunbound}
    t_\nu^{eq,unbound}(M)
    &\simeq
    \frac{8}{5}t_\gamma^{eq}(M)\\
     &+ \tau_{Hubble}\left(\frac{3}{5}
    \ln\left[
    \frac{m_\nu}{0.05\si{eV}}\right]+3.7
    \right)\nonumber
\end{flalign}

Equation \eqref{eq:tnuunbound} applies to black holes at rest with respect to the Hubble flow; however, black holes bound in galaxies and clusters typically have peculiar velocities of $v\simeq 10^{-3}c$. 
This quickly caps the Sommerfeld enchancement, and
\begin{flalign}
   t_\nu^{eq,bound}(M) 
   \simeq
   \frac{4}{3}t_\gamma^{eq}(M) \\
   &+ \tau_{Hubble}\left(\frac{1}{3}
    \ln\frac{m_\nu}{0.05\si{eV}} + 2.
   \right) \,.\nonumber
\end{flalign}
 
This additional delay is thus only by a  multiplicative factor of either $8/5$ or $4/3$, plus a small additional offset.
We thus expect to wait only 400-500 billion years until observed solar mass black holes begin evaporating.
Of course it will take about $10$ trillion years until the most massive black holes are emitting more Hawking radiation than they are absorbing CNB.

\section{The Cosmic Gravitational-Wave Background has little effect}
The cosmic gravitational-wave background (CGB) will not have much additional effect compared to photons, since we already know that the graviton mass is much much less than the CMB temperature, and so the energy density in the CGB is, and will long (or forever) remain, much less than that in the CMB.   

\section{Baryons and Dark Matter}

Baryons and dark matter have higher average energy densities than photons by a sizable factor,
after all
$\Omega_{matter}/\Omega_{\gamma} 
=(1+z_{eq}) \simeq 3400$.
This is much greater than $\Omega_\nu/\Omega_{\gamma} \simeq (4/11) (m_\nu/T_{CMB})=78~ (m_\nu/0.05\si{eV})$.

By how much black hole evaporation is delayed by baryon and dark matter infall depends on the relative velocities of the baryons and dark matter, and their local abundances in the vicinity of the black hole.
These are  environment-dependent, so
\begin{flalign}
    \label{eq:tmeqbound}
   t_m^{eq,bound}(M) 
   &\simeq
   \frac{4}{3}t_\gamma^{eq}(M)\\
   &+ \frac{1}{3}\tau_{Hubble}
   \left(\ln{[1+z_{eq}]} - \ln{\frac{v_{rel}}{c}} + \ln{\frac{\rho}{\Bar{\rho}}}\right)\nonumber\\
   \intertext{Inside structures (galaxies, clusters), $v_{rel} \gtrsim 10^{-3}c$, though outside them, it could fall much lower.
    Also average baryon and dark matter densities are up to $10^4$, or even further, enhanced relative to the cosmic average.
    Taking, conservatively, $v/c=10^{-3}$ and $\rho=\Bar{\rho}$
    }
    t_m^{eq,bound}(M) 
    &\geq
   \frac{4}{3}t_\gamma^{eq}(M) + 5.~ \tau_{Hubble}\,.
\end{flalign}

Similarly to neutrinos, this additional delay is only by a  multiplicative factor of $4/3$, plus a small  additional offset, though that is larger than that for neutrinos.
We thus still expect to wait only 400-500 billion years until observed solar-mass black holes begin evaporating.

\section{Stars}

There are sources of photons (and neutrinos and baryons) other than the cosmic backgrounds from the early universe.
Certainly here and now starlight, even excluding sunlight, impinges on us with greater power than the CMB.

After 400-500 billion years, when the CMB, CNB, and average matter flux are insufficient to keep solar mass black holes from evaporating, there will still be stars. 
Some of these will be stars that are currently extant.  
A $0.1M_\odot$ red dwarf has \cite{Adams:1996xe} an expected main sequence lifetime of $~10^{13}$y, and a luminosity of $10^{-3}{\cal L}_\odot\simeq4.\times10^{30}\si{erg/s}$.
Our galaxy has in excess of $10^9$ red dwarfs of this mass.
The distance at which the Hawking radiation from a black hole exceeds the intercepted luminosity from $N_{RD}$ of these red dwarfs is
\begin{equation}
    \label{eq:dRDeq}
    d_{RDeq} \simeq 5 \times 10^6\si{Mpc} \left(\frac{M}{M_\odot}\right)^2 \sqrt{N_{RD}}
\end{equation}
Thus solar or higher black holes cannot evaporate within the Hubble volume so long as there are any red dwarfs shining.

Star formation is an ongoing phenomena.  It has been argued \cite{Adams:1996xe} that, as galactic metallicities increase, star formation will favor lower mass stars, and possibly allow stars with masses as low as $0.04M_\odot$, with temperatures below $300$K, luminosities of $10^{-6}L_\odot$, and therefore lifetimes in excess of $10^{15}$y! 
For these future high metalicity, low mass red dwarfs
\begin{equation}
    \label{eq:dhighzRDeq}
    d_{highzRDeq} \simeq 2\times 10^5\si{Mpc} \left(\frac{M}{M_\odot}\right)^2 \sqrt{N_{high-z RD}}\,,
\end{equation}
still well beyond the Hubble distance.
We can therefore expect that, for at least the next $10^{15}\si{y}$, solar mass black holes will be prevented from evaporating and will not quantum spread.

\section{Black Holes in Galaxies and Clusters}

What happens to black holes inside galaxies and clusters in that very far distant time when all the stars are dark?
Even though those black holes will no longer be encountering starlight, 
the galaxies will still contain dark and ordinary matter.
That dark matter will continue to strike any black hole that passes through it.  For example, the local dark matter density is $\sim1\si{GeV/cm^3}$, and so, at a local velocity of $\sim 10^{-3}$c, a black hole would sweep up on average
\begin{equation}
    \frac{dM}{dt}\simeq 3\times10^{22}  \left(\frac{M}{M_\odot}\right)^2
    \si{GeV/s}
\end{equation}
as compared to radiating just
\begin{equation}
    {\cal L}_{Hawking} \simeq 4\pi R_s^2 \sigma_{SB} T_H^4 = 0.2 \left(\frac{M_\odot}{M}\right)^2 \si{GeV/s}\,.
\end{equation}
It thus remains impossible for a black hole to evaporate and quantum spread until the dark matter has been depleted.  
Adams and Laughlin \cite{Adams:1996xe} suggest that this will happen only in about $10^{25}$y.  

Should that long distant time come when black holes in galaxies are finally alone enough to evaporate, they are still bound to their galaxy.  
The classic calculation of their quantum spreading does not apply. 
(Actually, neither does the classic calculation of their Hawking radiation, which is  for isolated black holes.)

They may indeed begin to evaporate, but as for their quantum spreading, that will be much more limited than for an isolated black hole.
A black hole of mass $M$ in empty space moving at a velocity $\vec{v}\ll c$, and emitting a Hawking photon with energy $E$ in some direction $\hat{n}$, by conservation of momentum acquires velocity $\vec{v'}\simeq\vec{v}-(E/Mc)\hat{n}$.
Some time $t$ later its position differs from its original trajectory by $\Delta \vec{x}=-(E/Mc)\hat{n}t$.
This is the cause of the quantum spreading -- the black hole has an amplitude to emit that Hawking photon in every direction, and so, in the absence of measurement of either the black hole or the photon, its wavefunction spreads in position space.
On the other hand a black hole bound to a galaxy, and acquiring a small change in its velocity experiences only a small change in its energy, $\Delta E \simeq M\vec{v}\cdot\Delta\vec{v}\ll\frac{1}{2}M v^2$.  It thus remains bound to the galaxy.
The repeated Hawking photons are unsuccessful at causing the black hole to escaple until $T_H/M \simeq v_{esc}/c$.
This occurs only when $M\gtrsim M_{Pl}/\sqrt{8\pi v_{esc}/c}\simeq 10 M_{Pl}$.  
Thus until the final moments of Hawking radiation, the emission of the final $c/v_{esc}\lesssim 10^3$ Hawking photons, the black-hole wave function cannot spread beyond the size of the galaxy, or the size of the cluster if the black hole is outside a galaxy but in a cluster.  

Even that level of spreading would seem unlikely.  The black holes in galaxies will scatter \cite{Adams:1996xe} off of stars  approximately once in $10^{22}$y.
The spreading of the black hole in time $\Delta t$, can be expressed \cite{page_is_1980} as
\begin{flalign}
    \Delta x &= l_{Pl} \left(\frac{M_{Pl}}{M}\right)^{5/2} 
     \left(\frac{\Delta t}{t_{Pl}}\right)^{3/2}\\
    & \simeq 10^{-37} \si{pc} 
    \left(\frac{M_\odot}{M}\right)^{5/2} 
     \left(\frac{\Delta t}{10^{22}\si{y}}\right)^{3/2}
\end{flalign}
Thus scattering off of stars in the galaxy will prevent significant quantum spreading.
The black holes in the cluster but not bound to galaxies will spend a large enough fraction of their time in galaxies, that the same conclusion applies to them.

\section{Black Holes in intercluster space}

Finally, might we hope to observe black holes in intercluster space evaporate.
However, in a universe expanding exponentially with a Hubble time of just $\tau_{Hubble}$, it takes very little time for such a black hole starting a distance $x_i$ from bound object to be carried  beyond the anti-trapped surface at $r_{c}=c\tau_{Hubble}$ by the Hubble expansion,
\begin{equation}
    \Delta t = \tau_{Hubble}\ln\frac{c\tau_{Hubble}}{x_i}\,.
\end{equation}
For $x_i=100\si{Mpc}$, 
the logarithm is just $3.9$, 
giving $\Delta t \simeq 55\si{Gy}$,
for $x_i=1\si{kpc}$, it
gives $\Delta t \simeq 250\si{Gy}$.
For any object that is carried beyond $r_c$, we will never (so long as the current accelerated expansion continues) receive any signal that it emits after the crossing.  This is much like the situation for an object that passes through the horizon of a black hole -- we may continue to receive signals that it emitted before it crossed, but will never receive any signal it emits after it crossed.

Thus all black holes currently unbound to structures will exit the anti-trapped surface at $r_c$ and be lost to our view long before they can begin evaporating.
Black holes that are ejected from galaxies or clusters (evaporating in the classical sense)  in the far distant future have only a brief period compared to their Hubble lifetimes in which to undergo quantum spreading, before they too exit the anti-trapped surface.

\section{Conclusion}

We have shown that while isolated black holes are expected to evaporate and undergo quantum spreading of their wave functions, the real black holes that we currently observe -- i.e. black holes with masses greater than $M_\odot$ -- will not undergo such spreading for an extremely long time, or only do so beyond the Hubble horizon.
When the ones in our galaxy or cluster do eventually decay, the quantum spreading  will be severely constrained by the object being bound to the same  cosmic structures that we are, and look nothing like the quantum spreading of an isolated black hole.

\section*{Acknowledgements}QET and GDS were supported in part by a Department of Energy grant DE-SC0009946 to the particle astrophysics theory group at CWRU.

\bibliography{Bib}

\end{document}